\documentclass[12pt]{article}
\usepackage{a4}
\usepackage{amsmath}
\usepackage{amsfonts}
\usepackage{epsfig}

\input mydefs.cls

\begin{document}

\title{Asymptotic behavior of the ghost propagator in SU3 lattice gauge theory}

\date{\today}
\author{ Ph.~Boucaud$^a$, J.P.~Leroy$^a$, A.~Le~Yaouanc$^a$, A.Y.~Lokhov$^b$,\\
J. Micheli$^a$, O. P\`ene$^a$, J.~Rodr\'iguez-Quintero$^c$ and 
C.~Roiesnel$^b$ }
\par \maketitle

\begin{center}
$^a$Laboratoire de Physique Th\'eorique et Hautes
Energies\footnote{Unit\'e Mixte de Recherche 8627 du Centre National de 
la Recherche Scientifique}\\
{Universit\'e de Paris XI, B\^atiment 211, 91405 Orsay Cedex,
France}\\
$^b$ Centre de Physique Th\'eorique\footnote{
Unit\'e Mixte de Recherche C7644 du Centre National de 
la Recherche Scientifique\\ 
}de l'Ecole Polytechnique\\
F91128 Palaiseau cedex, France\\ 
$^c$ Dpto. F\'isica Aplicada, Fac. Ciencias Experimentales,\\
Universidad de Huelva, 21071 Huelva, Spain.
\end{center}

\begin{abstract}

  We study the asymptotic behavior of the ghost propagator in the
  quenched SU(3) lattice gauge theory with Wilson action. The study is
  performed on lattices with a physical volume fixed around 1.6~fm and
  different lattice spacings: 0.100~fm, 0.070~fm and 0.055~fm. We
  implement an efficient algorithm for computing the Faddeev-Popov
  operator on the lattice. We are able to extrapolate the lattice data
  for the ghost propagator towards the continuum and to show that the
  extrapolated data on each lattice can be described up to four-loop
  perturbation theory from 2.0 GeV to 6.0 GeV. The three-loop values
  are consistent with those extracted from previous perturbative
  studies of the gluon propagator. However the effective
  $\Lambda_{\ms}$ scale which reproduces the data does depend strongly
  upon the order of perturbation theory and on the renormalization
  scheme used in the parametrization. We show how the truncation of
  the perturbative series can account for the magnitude of the
  dependency in this energy range. The contribution of
  non-perturbative corrections will be discussed elsewhere.

\end{abstract}

\begin{flushleft}
LPT-Orsay 05-37\\
CPHT RR 037.0605\\
UHU-FT/05-11
\end{flushleft}

\enlargethispage{0.5cm}

\newpage
\section{Introduction}

Whereas lattice gauge theory (LGT) has been initially formulated to
study gauge-invariant quantities in the non-perturbative regime, it
has long been recognized that LGT could be a useful tool for studying
gauge-variant quantities such as Green functions, both in the
non-perturbative and in the perturbative regimes. The SU(3) gluon
propagator in momentum space was first considered \cite{BER94} to gain
some insight into the physics of confinement.  Much work was then
devoted to the study of its infrared behavior (for a review see
\cite{MAN99}).  Subsequent studies \cite{LPTX99,LPTX00} were focused on the
ultraviolet behavior and have been able to compare quantitatively the
large momentum dependence of the lattice gluon propagator with
perturbative predictions beyond one-loop order. The result for
$\Lambda_{\ms}$ was found to depend strongly upon the order of the
perturbation theory and upon the renormalisation scheme used in the
parametrization. This strong dependence raised the question whether
the energy windows in these calculations were large enough for
perturbative QCD to be a valid approximation.

On the other hand, as shown by Gribov \cite{GRI78}, the infrared
behavior of the gluon propagator is closely related to the singularity
structure of the ghost propagator inferred from the gauge-fixing
ambiguities. As is well-known, the Landau gauge, which is presently
the only covariant gauge for which there exists effective local
gauge-fixing algorithms on the lattice, suffers from these
ambiguities. The comprehensive theoretical study by Zwanziger
\cite{ZWA94} of the Faddeev-Popov operator on the lattice in Landau
gauge spurred the first numerical study of the ghost propagator
\cite{SUM96} in SU(2) and SU(3) gauge theories. Most subsequent
activity has been dedicated to the SU(2) lattice gauge theory in the
infrared region, mainly for technical reasons as we shall explain
below. There are relatively few numerical studies of the SU(3) ghost
propagator which are either more focused on the infrared region and
the Gribov copy problem \cite{STE05a,STE05b,STE05c} or have only
performed a qualitative perturbative description in the quenched
approximation \cite{JAP04a,JAP04b} and, quite recently, in the
unquenched case also \cite{JAP05}.

It is important to make the study of the SU(3) ghost propagator in the
ultraviolet region more quantitative for comparison purposes with the
gluon propagator. Lattice results at small distances may be described
by perturbation theory and the independent extraction of the
$\Lambda_{\text{QCD}}$ scale from the two propagators would provide a
self-consistency test of the analysis and of the lattice approach. It
would be particularly significant to confirm or not, from the study of
the lattice propagators alone, the need for the non-perturbative power
corrections found in the study of the three-gluon coupling on the
lattice \cite{BOU00}.

The paper is organized as follows. We will begin by recalling in
section~\ref{sec:charge} the method used to relate lattice data for
the ghost propagator to its perturbative renormalization description.
Then we proceed by exhibiting in section~\ref{sec:lattice} the salient
features of our lattice calculation, particularly of our
implementation of the Faddeev-Popov operator on the lattice. The
following section outlines the general method that we devised
previously \cite{LPTX99,LPTX00} to eliminate hypercubic artifacts from
two-point functions and extrapolate the lattice data towards the
continuum. This extrapolation is crucial to succeed in a quantitative
description. The results are discussed in section~\ref{sec:analysis}
which contains several subsections where the analysis is performed in
different renormalization schemes up to four-loop order. In particular
the scheme dependence is thoroughly investigated and used to probe the
effects of the truncation of the perturbative series. We conclude in
section \ref{sec:conclusion} with a comparison of the different
methods to compute the $\Lambda_{\text{QCD}}$ scale on the lattice.

\section{Renormalization description of the ghost propagator}
\label{sec:charge}

Let $\Gamma^{(n)}_{B}$ be some gauge-fixed multiplicatively
renormalizable one-particle irreducible $n$-point bare Green function
defined in euclidean momentum space and in some regularization scheme
with cut-off $\Lambda$. Let $s$ denotes some polarization state and
kinematical configuration of the external particles contributing to
$\Gamma^{(n)}_{B}$. Let $p$ denote a scale transformation on $s$ and
$g_{B}$ denote the bare coupling. It is well known that, in any
renormalization scheme $R$ defined by some renormalization conditions
on state $s$ at the renormalization point $p=\mu$, we have
\begin{eqnarray}
  \label{eq:renormalization}
  \Gamma^{(n)}_{B}(p,s,g_{B},\Lambda) = Z_{\Gamma,R}(\mu,s,g_{R},\Lambda)
  \Gamma^{(n)}_{R}(p,s,g_{R},\mu) + {\cal O}(\Lambda^{-1})
\end{eqnarray}
where $Z_{\Gamma,R}$ is the renormalization constant in scheme $R$,
$\Gamma^{(n)}_{R}$ is the renormalized Green function and $g_{R}(\mu)$
is the renormalized coupling. We omit the dependence on the gauge
parameter for simplicity of notation since we will specialize to
Landau gauge.

The explicit dependence on $\mu$ drops out of the renormalized Green
function $\Gamma^{(n)}_{R}$ at the renormalization point $p=\mu$. It
follows that
\begin{align}
  \label{eq:evolution}
  \begin{split}
  \lim_{\Lambda\rightarrow\infty}
  \frac{d\ln\left(\Gamma^{(n)}_{B}(\mu,s,g_{B},\Lambda)\right)}
  {d\ln \mu^{2}} &=
  \lim_{\Lambda\rightarrow\infty}
  \frac{d\ln\left(Z_{\Gamma,R}(\mu,s,g_{R},\Lambda)\right)}
  {d\ln \mu^{2}} + \frac{d\ln\left(\Gamma^{n}_{R}(s,g_{R})\right)}
  {d\ln \mu^{2}} \\
   &\equiv \gamma_{\Gamma,R}(g_{R}) 
   + \frac{d g_{R}}{d \ln\mu^{2}}\frac{\partial\ln\Gamma^{n}_{R}}{\partial g_{R}}
 \end{split}   
\end{align}
The arbitrariness in the choice of the renormalization scheme $R$ has
prompted attempts at determining the ``best'' schemes for describing
the $q^{2}$-evolution of bare Green functions on the lattice. Clearly
it is always possible to find a change of coupling which will be a
best approximation of a set of data at a given order of perturbation
theory, within some prescribed criteria. Rather than pursuing this
route, we will follow the standard wisdom which consists in choosing
renormalization conditions appropriate to the continuum quantity under
scrutiny.

Momentum substraction schemes have long been used to define
renormalization conditions befitted to the description of the
renormalization dependence of ``physical'' quantities. They are
defined by setting some of the 2 and 3-point functions to their tree
values.  In the $\mom$ schemes, for these Green functions,
Eq.\,(\ref{eq:evolution}) simplifies to
\begin{align}
  \label{eq:mom}
  \lim_{\Lambda\rightarrow\infty}
  \frac{d\ln\left(\Gamma^{(n)}_{B}(\mu,s,g_{B},\Lambda)\right)}
  {d\ln \mu^{2}} &= \frac{d\ln(Z_{\Gamma,MOM})}{d \ln \mu^{2}} =
  \gamma_{\Gamma,MOM}(g_{MOM})
\end{align}
Infinitely many MOM schemes can be defined which differ by the
substraction point of the vertices. We have shown in \cite{LPTX99} that
the $\momg$ scheme defined by substracting the transversal part of the
three-gluon vertex at the asymmetric point where one external momentum
vanishes, appears to provide a much better estimate of the asymptotic
behavior of the gluon propagator in Landau gauge than the $\ms$
scheme. For the study of the asymptotic behavior of the ghost
propagator in Landau gauge, it seems therefore natural to use a
$\momc$ scheme defined by substracting the ghost-gluon vertex at the
asymmetric point where the momentum of the external gluon vanishes.
Comparison of the two $\mom$ schemes should provide us with an estimate
of the systematic error entailed in the truncation of the perturbation
theory.

\begin{figure}[h]
  \centering
  \psfig{figure=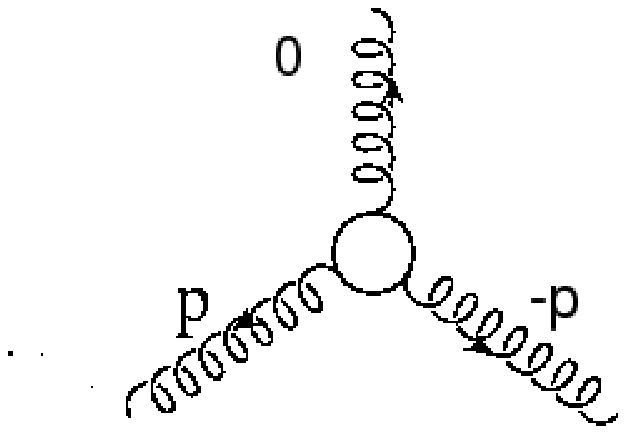, width=4cm, height=4cm}
  \hskip 2cm
  \psfig{figure=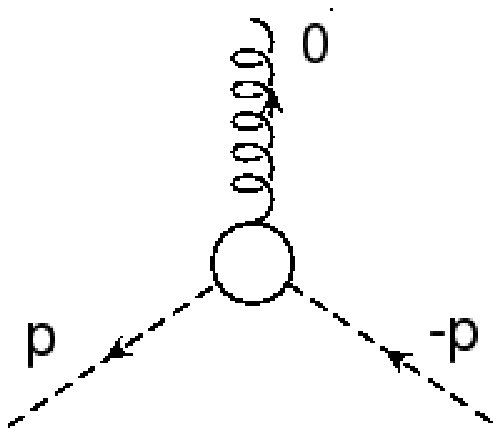, width=4cm, height=4cm}
  \caption{$\momg$ scheme (left) and $\momc$ scheme (right).}
  \label{fig:mom}
\end{figure}

The perturbative calculation of the gluon, ghost and quark
self-energies and all 3-vertices appearing in the QCD Lagrangian have
been done at three-loop order in the $\ms$ scheme and in a general
covariant gauge at the asymmetric point with one vanishing momentum
\cite{CHE00}. These three-loop results allow one to relate the coupling
constants of any $\mom$-like scheme to the $\ms$ scheme at three-loop
order. For the $\momg$ and $\momc$ schemes defined above these
relations read respectively in Landau gauge and in the quenched
approximation ($n_{f}=0$), with $\displaystyle{h =
  \frac{g^{2}}{16\pi^{2}}}$:
\begin{align}
  \label{eq:momg}
  \begin{split}
  h_{\momg} &= h_{\ms} + \frac{70}{3}\,h^{2}_{\ms}
            + \left(\frac{516217}{576}-\frac{153}{4}\zeta_{3}\right)h^{3}_{\ms}
 \\ 
            &+ \left(\frac{304676635}{6912} - \frac{299961}{64}\zeta_{3} -
                    \frac{81825}{64}\zeta_{5} \right)h^{4}_{\ms}
  \end{split}
  \\
  \label{eq:momc}
  \begin{split}
  h_{\momc} &= h_{\ms} + \frac{223}{12}\,h^{2}_{\ms}
            + \left(\frac{918819}{1296}-\frac{351}{8}\zeta_{3}\right)h^{3}_{\ms}
 \\ 
            &+ \left(\frac{29551181}{864} - \frac{137199}{32}\zeta_{3} -
                    \frac{74295}{64}\zeta_{5} \right)h^{4}_{\ms}
  \end{split}
\end{align}
The very large coefficients of these perturbative expansions explain
the difficulties met by the $\ms$ scheme to approach asymptotic
scaling below 10 GeV.

The recent calculation \cite{CHE05} of the anomalous dimensions in the
$\ms$ scheme of the gluon and ghost fields at four-loop order,
together with the knowledge of the $\beta$-function \cite{RIT97},
makes it possible to perform the analysis of the lattice data for the
gluon and ghost propagators up to four-loop order also in the $\momg$
and $\momc$ schemes. The numerical coefficients of the
$\beta$-function defined as
\begin{align}
  \label{eq:beta_f}
  \beta(h) = \frac{dh}{d\ln \mu^{2}} = - \sum_{i=0}^{n} \beta_{i}h^{i+2} +
  {\cal O}(h^{n+3}) 
\end{align}
are:
\begin{align}
  \label{eq:coeffs}
  \beta_{2}^{\momg} = 2412.16,\ \ \beta_{2}^{\momc} = 2952.73,
  \ \beta_{3}^{\momg} = 84353.8,\ \beta_{3}^{\momc} = 101484.
\end{align}

For completeness we also give the expansion coefficients of the
renormalisation constants of the gluon and ghost fields in the MOM
schemes with respect to the renormalized coupling of the $\ms$ scheme
up to four-loop order:
\begin{align}
  \label{eq:Z3}
  \begin{split}
 \frac{d\ln(Z_{3,MOM})}{d \ln \mu^{2}} &= 
 \frac{13}{2}\,h_{\ms} + \frac{3727}{24}\,h^{2}_{\ms} + 
 \left(\frac{2127823}{288} - \frac{9747}{16}\zeta_{3}\right) h^{3}_{\ms} \\
 &+ \left(\frac{3011547563}{6912} - \frac{18987543}{256}\zeta_{3} - 
   \frac{1431945}{64}\zeta_{5}\right) h^{4}_{\ms} 
  \end{split}
  \\
  \label{eq:Z3c}
  \begin{split}
 \frac{d\ln(\widetilde{Z}_{3,MOM})}{d \ln \mu^{2}} &= 
 \frac{9}{4}\,h_{\ms} + \frac{813}{16}\,h^{2}_{\ms} + 
 \left(\frac{157303}{64} - \frac{5697}{32}\zeta_{3}\right) h^{3}_{\ms} \\
 &+ \left(\frac{219384137}{1536} - \frac{9207729}{512}\zeta_{3} - 
   \frac{221535}{32}\zeta_{5}\right) h^{4}_{\ms} 
  \end{split}
\end{align}
\section{Lattice calculation}
\label{sec:lattice}

\subsection{Faddeev-Popov operator on the lattice}
\label{sec:FP}

The ghost propagator is defined on the lattice as
\begin{align}
  \label{eq:Ghost}
  G(x-y)\delta^{ab} \equiv \left<\left(M^{-1}\right)^{ab}_{xy}\right>
\end{align}
where the action of the Faddeev-Popov operator $M$ on an arbitrary
element $\eta$ of the Lie algebra ${\cal SU}$(N) of the gauge group
SU(N), in a Landau gauge fixed configuration, is given by
\cite{ZWA94}:
\begin{align}
\label{eq:FP1}
\nonumber
(M\eta)^{a}(x) &= \frac{1}{N}\sum_{\mu}\biggl\{ G_{\mu}^{ab}(x)
          \left(\eta^{b}(x+\hat{\mu})-\eta^{b}(x)\right)
        - (x \leftrightarrow x-\hat{\mu}) \\
 &\qquad\qquad + \frac{1}{2}f^{abc}(
             \eta^{b}(x+\hat{\mu})A_{\mu}^{c}(x)
           - \eta^{b}(x-\hat{\mu})A_{\mu}^{c}(x-\hat{\mu}) \bigr)
          \biggr\}
\end{align}
and where, with antihermitian generators $T^{a}$,
\begin{align}
\label{eq:defa}
G_{\mu}^{ab}(x) &= \frac{1}{2}\tr\left(\left\{T^{a},T^{b}\right\}
\left(U_{\mu}(x)+U_{\mu}^{\dagger}(x)\right)\right) \\
\label{eq:defb}
A_{\mu}^{c}(x) &= 
-\tr\left(T^{c}\left(U_{\mu}(x)-U_{\mu}^{\dagger}(x)\right)\right)
\end{align}
Most lattice implementations of the Faddeev-Popov operator have
followed closely the component-wise Eqs.\,(\ref{eq:FP1}-\ref{eq:defb}).
But the derivation in \cite{ZWA94} shows that the
Faddev-Popov operator can also be written as a lattice divergence:
\begin{align}
\label{eq:FP2a}
M(U) = -\frac{1}{N} \nabla\cdot \widetilde{D}(U)
\end{align}
where the operator $\widetilde{D}$ reads
\begin{align}
\label{eq:FP2b}
\widetilde{D}_{\mu}(U)\eta(x) 
   &= \frac{1}{2}\left(U_{\mu}(x)\eta(x+\hat{\mu}) -\eta(x)U_{\mu}(x)
    + \eta(x+\hat{\mu})U^{\dagger}_{\mu}(x)
    - U^{\dagger}_{\mu}(x)\eta(x) \right)
\end{align}
Using conversion routines between the Lie algebra and the Lie group,
eqs.~(\ref{eq:FP2a}-\ref{eq:FP2b}) allow for a very efficient lattice
implementation, sketched in Table~\ref{tab:algo}, which is based
on the fast routines coding the group multiplication law.
\begin{table}[h]
\center{\rule{10cm}{1pt}}
\begin{align*}
&!\ \eta_{in},\ \eta_{out}\ \text{are the ghost fields.} \\
&!\ U\ \text{is the gauge configuration.} \\ 
&\mathrm{type~(SUN)}\quad U(*), dU, W, W_{+}, W_{-} \\
&\mathrm{type~({\cal SU}N)}\quad \eta_{in}(*), \eta_{out}(*) \\
&\mathrm{for~all~x:}\\
&\quad dU = 0. \\
&\quad W = \eta_{in}(x) \\ 
&\quad \mathrm{do}\ \mu = 1,\ 4 \\
   &\quad\qquad W_{+} = \eta_{in}(x+\hat{\mu}) \\ 
   &\quad\qquad W_{-} = \eta_{in}(x-\hat{\mu}) \\ 
   &\quad\qquad dU = dU + U_{\mu}(x-\hat{\mu})\times W + W\times U_{\mu}(x) \\
   &\quad\qquad\quad\qquad\quad - U_{\mu}(x)\times W_{+} 
   - W_{-}\times U_{\mu}(x-\hat{\mu}) \\
&\quad \mathrm{enddo} \\
&\quad \eta_{out}(x) = dU - dU^{\dagger} - \frac{1}{N}\tr(dU- dU^{\dagger})
\end{align*}
\center{\rule{10cm}{1pt}}
\caption{Pseudo code of our implementation of the Faddeev-Popov operator.}
\label{tab:algo}
\end{table}

\subsection{Inversion of the Faddeev-Popov operator}
\label{sec:inversion}

Constant fields are zero modes of the Faddeev-Popov operator. This
operator can be inverted only in the vector subspace $K^{\perp}$
orthogonal to its kernel. If the Faddeev-Popov operator has no other
zero modes than constant fields, then the non-zero Fourier modes form
a basis of $K^{\perp}$:
\begin{align}
  \label{eq:ortho}
  \eta(x) = \sum_{p\neq 0} c_{p}e^{ip\cdot x}\,,\quad \forall \eta\in K^{\perp}
\end{align}
The standard procedure has been to invert the Faddev-Popov operator
with one non-zero Fourier mode as a source
\begin{align}
   S^{a}_{p}(x) = \delta^{ab}e^{ip\cdot x}
\end{align}
and to take the scalar product of $M^{-1}S^{a}_{p}$ with the source:
\begin{align}
  \left(S^{a}_{p}\left|\right.M^{-1}S^{a}_{p}\right) &= 
  \sum_{x,y}\left(M^{-1}\right)^{aa}_{xy}e^{-ip\cdot(x-y)} \\
  \label{eq:ftp}
  &= V\,\widehat{G}(p)
\end{align}
after averaging over the gauge field configurations.  This method
requires one matrix inversion for each value of the ghost propagator
in momentum space. It is suitable only when one is interested in a few
values of the ghost propagator.

However, the study of the ultraviolet behavior of the ghost propagator
in the continuum requires its calculation at many lattice momenta to
control the spacing artifacts, as we shall see in the next section.
This can be done very economically by noting that
\begin{align}
  \label{eq:delta}
  \delta(x,y) = \frac{1}{V} + \frac{1}{V}\sum_{p\ne 0} e^{-ip\cdot(x-y)}
\end{align}
and choosing as source:
\begin{align}
  \label{eq:zero}
  S^{a}_{0}(x) = \delta^{ab}\left(\delta(x,0) - \frac{1}{V}\right)
\end{align}
The Fourier transform of $M^{-1}S^{a}_{0}$, averaged over the gauge
configurations, yields:
\begin{align}
  \nonumber
  \sum_{x} e^{-ip\cdot x}\left<M^{-1}S^{a}_{0}\right> &=
  \sum_{x} e^{-ip\cdot x}\left<\left(M^{-1}\right)^{aa}_{x0}\right> - 
  \frac{1}{V}\sum_{x,y}e^{-ip\cdot x}
  \left<\left(M^{-1}\right)^{aa}_{xy}\right> \\
  \nonumber
  &= \sum_{x} e^{-ip\cdot x}G(x) -
  \frac{1}{V}\sum_{x,y}e^{-ip\cdot x}G(x-y) \\
  &=  \widehat{G}(p) - \delta(p)\sum_{x}G(x) 
  \label{eq:ft0}
\end{align}
as a consequence of the translation invariance of the ghost
propagator.  Therefore, with this choice of source, only one matrix
inversion followed by one Fourier transformation of the solution is
required to get the full ghost propagator on the lattice.

There is of course a price to pay, as can be read off
Eq.\,(\ref{eq:ft0}) which lacks the factor $V$ present in
Eq.\,(\ref{eq:ftp}). The statistical accuracy with the source
$S^{a}_{p}$ is better, especially at high momentum $p$. However the
statistical accuracy with the source $S^{a}_{0}$ turns out to be
sufficient for our purpose.

There is one final point we want to make and which has never beeen
raised to the best of our knowledge. It is mandatory to check,
whatever the choice of sources, that rounding errors during the
inversion do not destroy the condition that the solution belongs to
$K^{\perp}$:
\begin{align}
  \label{eq:kernel}
  \sum_{x}\left(M^{-1}S\right)(x) = 0
\end{align}
Indeed, if the zero-mode component of the solution grows beyond some
threshold during the inversion of the Faddeev-Popov operator on some
gauge configuration, then that component starts to increase
exponentially and a sizeable bias is produced in other components as
well. We have observed this phenomenon occasionally, about one gauge
configuration every few hundreds, when using the implementation of the
lattice Faddeev-Popov operator based on
Eqs.\,(\ref{eq:FP1}-\ref{eq:defb}).  But the systematic bias which is
induced on the averages over gauge field configurations can be
uncomfortably close to those ascribed to Gribov copies.

Another virtue of the algorithm described in Table~\ref{tab:algo} is
its numerical stability which is improved by several orders of
magnitude. We have never observed sizeable deviations from
Eq.\,(\ref{eq:kernel}) with this algorithm.

\subsection{The simulation}
\label{sec:simulation}

We ran simulations of the $SU(3)$ lattice gauge theory with the Wilson
action in the quenched approximation on several hypercubic lattices,
whose parameters are summarized in Table~\ref{tab:simulation}. All
lattices have roughly the same physical volume except the $24^{4}$
lattice at $\beta=6.0$ which has been included to check out
finite-volume effects.
\begin{table}
  \centering
\begin{tabular}[h]{cccc}
  \hline
  $\beta$ & $V$ &  $a^{-1}$ (GeV) &\# Configurations \\
  \hline
  $6.0$   &  $16^{4}$ & $1.96$ & $1000$ \\
  $6.0$   &  $24^{4}$ & $1.96$ & $500$ \\
  $6.2$   &  $24^{4}$ & $2.75$ & $500$  \\
  $6.4$   &  $32^{4}$ & $3.66$ & $250$ \\
  \hline
\end{tabular}
\caption{Run parameters. The lattice spacings are taken from Table~3 in \cite{BAL93} with a physical unit normalized by $\sqrt{\sigma}=445$ MeV.}
\label{tab:simulation}
\end{table}
The SU(3) gauge configurations were generated using a hybrid algorithm
of Cabibbo-Marinari heatbath and Creutz overrelaxation steps. 10000
lattice updates were discarded for thermalization and the
configurations were analyzed every 100/200/500 sweeps on the
$16^{4}/24^{4}/32^{4}$ lattices.

Landau gauge fixing was carried out by minimizing the functional
\begin{align}
  \label{eq:landau}
  F_{U}[g] =\text{Re}\sum_{x}\sum_{\mu}
  \left(1-\frac{1}{N}g(x)U_{\mu}(x)g^{\dagger}(x+\hat{\mu})\right)
\end{align}
by use of a standard overrelaxation algorithm driving the gauge
configuration to a local minimum of $F_{U}[g]$. We did not try to
reach the fundamental modular region $\Lambda$, defined as the set of
absolute minima of $F_{U}[g]$ on all gauge orbits. Indeed there have
been numerous studies, in SU(2) \cite{CUC97,BAK04} and in SU(3)
\cite{STE05a,STE05b}, of the effect of Gribov copies on the ghost
propagator.  The consensus is that noticeable systematic errors,
beyond statistical errors, are only found for the smallest $p^{2}$,
much smaller than the squared momenta that we used to study the
asymptotic behavior of the ghost propagator.

Then the ghost propagator $G(p)$ is extracted from Eq.\,(\ref{eq:ft0})
for all $p\neq 0$. The required matrix inversion, with a
conjugate-gradient algorithm without any preconditioning, and the
Fourier transform consume in average less than half the computing time
of the Landau gauge fixing.

\section{Hypercubic artifacts}
\label{sec:artifact}
The ghost propagator $\widehat{G}(p)$ is a scalar invariant on the
lattice which means that it is invariant along the orbit $O(p)$
generated by the action of the isometry group $H(4)$ of hypercubic
lattices on the discrete momentum
$p\equiv\frac{2\pi}{La}\times(n_{1},n_{2},n_{3},n_{4})$ where the
$n_{\mu}$'s are integers, $L$ is the lattice size and $a$ the lattice
spacing.  The general structure of polynomials invariant under a
finite group is known from group-invariant theory. Indeed it can be shown
that any polynomial function of $p$ which is invariant under the
action of $H(4)$ is a polynomial function of the 4 invariants $p^{[n]}
= a^{n}\sum_{\mu}p_{\mu}^{n}, n = 2, 4, 6, 8$ which index the set of
orbits.

Our analysis program uses these 4 invariants to average the ghost
propagator over the orbits of $H(4)$ to increase the statistical
accuracy:
\begin{align}
  a^{2}G_{L}(p^{[2]},p^{[4]},p^{[6]},p^{[8]}) =
  \frac{1}{\|O(p)\|} \sum_{p\in O(p)} \widehat{G}(p)
\end{align}
where $\|O(p)\|$ is the cardinal number of the orbit $O(p)$. By the
same token, one should always take the following \bfit{real} source
\begin{align}
  \label{eq:source}
  \overline{S}^{a}_{p}(x) = \delta^{ab}\sum_{p\in O(p)} \cos(p\cdot x)
\end{align}
rather than a single complex Fourier mode for studies of the ghost
propagator in the infrared region. Indeed, after averaging over the
gauge configurations and use of the translational invariance, one gets
\begin{align}
  \nonumber
  \left<\left(
      \overline{S}^{a}_{p}\left|\right.M^{-1}\overline{S}^{a}_{p}
    \right)\right> &= 
  \sum_{p,p'\in O(p)}\sum_{x,y}\left<\left(M^{-1}\right)^{aa}_{xy}\right>
  e^{-ip'\cdot x+ip\cdot y} \\
  \label{eq:fto}
  &= V\|O(p)\|\,a^{2}G_{L}(p^{[2]},p^{[4]},p^{[6]},p^{[8]})
\end{align}
By analogy with the free lattice propagator
\begin{align}
  \label{eq:free}
  G_{0}(p) = \frac{1}{\sum_{\mu}\widehat{p}_{\mu}^2} =
  \frac{a^{2}}{p^{[2]}}\left(1+\frac{1}{12}\frac{p^{[4]}}{p^{[2]}} +
    \cdots\right)\,, \quad\mathrm{where}\quad\widehat{p}_{\mu} =
  \frac{2}{a}\sin\left(\frac{ap_{\mu}}{2}\right)
\end{align}
it is natural to make the hypothesis that the lattice ghost propagator
is a smooth function of the discrete invariants near the continuum
limit, when $a\,p_{\mu}\ll 1\,,\forall\mu$,
\begin{align}
  \label{eq:invariants}
  G_{L}(p^{[2]},p^{[4]},p^{[6]},p^{[8]}) \approx 
  G_{L}(p^{[2]},0,0,0) + p^{[4]}\frac{\partial G_{L}}
  {\partial p^{[4]}}(p^{[2]},0,0,0) + \cdots
\end{align}
and $G_{L}(p^{[2]},0,0,0)$ is nothing but the propagator of the
continuuum in a finite volume, up to lattice artifacts which do not
break $O(4)$ invariance. When several orbits exist with the same
$p^{2}$, the simplest method to reduce the hypercubic artifacts is to
extrapolate the lattice data towards $G_{L}(p^{[2]},0,0,0)$ by making
a linear regression at fixed $p^{2}$ with respect to the invariant
$p^{[4]}$ since the other invariants are of higher order in the
lattice spacing. The range of validity of this linear approximation
can be checked a posteriori from the smoothness of the extrapolated
data with respect to $p^{2}$.

Choosing the variables $\widehat{p}_{\mu}$ appropriate to the
parametrization of a lattice propagator with periodic boundary
conditions provides an independent check of the extrapolation. Indeed we
can write as well
\begin{align}
  G_{L}(p^{[2]},p^{[4]},p^{[6]},p^{[8]}) \equiv
  \widehat{G}_{L}(\widehat{p}^{[2]},\widehat{p}^{[4]},\widehat{p}^{[6]},
  \widehat{p}^{[8]})
\end{align}
with the new invariants, again hierachically suppressed with respect
to the lattice spacing,
\begin{align}
  \widehat{p}^{[n]} = a^{n}\sum_{\mu}\widehat{p}_{\mu}^{n} 
\end{align}
$G_{L}$ and $\widehat{G}_{L}$ are two different parametrizations of
the same lattice data, but near the continuum limit one must also have
\begin{align}
  \label{eq:sinus}
  \widehat{G}_{L}(\widehat{p}^{[2]},\widehat{p}^{[4]},\widehat{p}^{[6]},
          \widehat{p}^{[8]}) \approx 
  \widehat{G}_{L}(\widehat{p}^{[2]},0,0,0) + \widehat{p}^{[4]}
\frac{\partial \widehat{G}_{L}}
  {\partial \widehat{p}^{[4]}}(\widehat{p}^{[2]},0,0,0) + \cdots  
\end{align}
where $G_{L}(p^{[2]},0,0,0)$ and
$\widehat{G}_{L}(\widehat{p}^{[2]},0,0,0)$ are the \bfit{same}
function, the propagator of the continuum , of a \bfit{different}
variable (again up to lattice artifacts which do not break $O(4)$
invariance).

If one wants to include in the data analysis the points with a single
orbit at fixed $p^{2}$, one must interpolate the slopes extracted from
Eqs~(\ref{eq:invariants}) or (\ref{eq:sinus}). This interpolation can
be done either numerically or by assuming a functional dependence of
the slope with respect to $p^{2}$ based on dimensional arguments. The
simplest ansatz is to assume that the slope has the same leading
behavior as for a free lattice propagator:
\begin{align}
  \label{eq:slope}
  \frac{\partial G_{L}}
  {\partial p^{[4]}}(p^{[2]},0,0,0) &=
 \frac{1}{\left(p^{[2]}\right)^{2}}\left( c_{1}+ c_{2}p^{[2]}\right)
\end{align}
The inclusion of $O(4)$-invariant lattice spacing corrections is
required to get fits with a reasonable $\chi^{2}$. The quality of such
two-parameter fits to the slopes, and the extension of the fitting
window in $p^{2}$, supplies still another independent check of the
validity of the extrapolations.
 
We have used Eqs.~(\ref{eq:invariants}) and (\ref{eq:slope}) to
extrapolate our lattice data towards the continuum and determined the
range of validity in $p^{2}$ of the extrapolations from the
consistency of the different checks within our statistical errors.
The errors on the extrapolated points have been computed with the
jackknife method. Tables~\ref{tab:cutg} and \ref{tab:cutc} summarize
the cuts that have been applied to the data for the estimation of the
systematic errors in the analysis of the next section. We have
repeated the analysis of the gluon propagator \cite{LPTX00} to study
the sensitivity of the results with respect to the window in $p^{2}$
which has been enlarged considerably in our new data. The cuts for the
lattice ghost propagator are stronger than for the gluon lattice
propagator because the statistical errors of the former are two to
three times larger which make the continuum extrapolations less
controllable.

\begin{table}[h]
\centering
\begin{tabular}{c|c|c|c|c|c}
  \hline
  $\beta$ & $V$ & $N_{points}$ &  $a\,p_{min}$ & $a\,p_{max}$ & $\chi^{2}$ \\
  \hline
  $6.0$   &  $16^{4}$ & $> 10$ & $\leq 1.30$ & $\leq 1.82$ &  $\leq 1.4$ \\
  $6.2$   &  $24^{4}$ & $> 12$ & $\leq 1.30$ & $\leq 1.82$ &  $\leq 1.1$  \\
  $6.4$   &  $32^{4}$ & $> 20$ & $\leq 1.40$ & $\leq 1.82$ &  $\leq 1.3$ \\
\hline
\end{tabular}
\caption{Cuts on the lattice data for the gluon propagator. 
$[a\,p_{min},a\,p_{max}]$ is the momentum window of a fit in lattice units and 
$N_{points}$ is the number of data points in that window.}
\label{tab:cutg}
\end{table}

\begin{table}[h]
\centering
\begin{tabular}{c|c|c|c|c|c}
  \hline
  $\beta$ & $V$ & $N_{points}$ &  $a\,p_{min}$ & $a\,p_{max}$ & $\chi^{2}$ \\
  \hline
  $6.0$   &  $16^{4}$ & $> 10$ & $\leq 1.30$ & $\leq 1.57$ & $\leq 1.0$ \\
  $6.2$   &  $24^{4}$ & $> 20$ & $\leq 1.30$ & $\leq 1.57$ & $\leq 1.0$  \\
  $6.4$   &  $32^{4}$ & $> 20$ & $\leq 1.00$ & $\leq 1.57$ & $\leq 1.0$  \\
\hline
\end{tabular}
\caption{Cuts on the lattice data for the ghost propagator. 
Columns have the same meaning as in Table\,\ref{tab:cutg}.}
\label{tab:cutc}
\end{table}

The number of distinct orbits at each $p^{2}$ increases with the
lattice size and, eventually, a linear extrapolation limited to the
single invariant $p^{[4]}$ breaks down.  However there is a systematic
way to include higher-order invariants and to extend the range of
validity of the extrapolations. A much more detailed exposition of the
controlling of systematic errors is in preparation, since our method
has been largely ignored in the litterature where very empirical
recipes are still in use.

\section{Data analysis}
\label{sec:analysis}

The evolution equation of the renormalization constants of the gluon
or ghost fields in a MOM scheme, with respect to the coupling constant
$h$ in an arbitrary scheme $R$ (the index $R$ is omitted but
understood everywhere), can be written generically up to four-loop
order:
\begin{align}
  \label{eq:dZmom}
\frac{d\ln(Z_{\Gamma,MOM})}{d \ln \mu^{2}} =
\overline{\gamma}_{0} h
+ \overline{\gamma}_{1} h^{2} + \overline{\gamma}_{2} h^{3} 
+ \overline{\gamma}_{3} h^{4} 
\end{align}
and the perturbative integration of Eq.~(\ref{eq:dZmom}) yields, to the
same order,
\begin{align}
  \label{eq:Zmom}
  \begin{split}
     \ln\left(\frac{Z_{\Gamma,MOM}}{Z_{0}}\right) & = 
     \log(h)\,\frac{\overline{\gamma}_{0}}{\beta_{0}} + 
     h\,\frac{\left(\beta_{0}\,\overline{\gamma}_{1}-\beta_{1}\,\overline{\gamma}_{0}\right)}{\beta_{0}^{2}} \\
     &+ h^{2}\,\frac{\left(\beta_{0}^{2}\,\overline{\gamma}_{2}-\beta_{0}\,\beta_{1}\,\overline{\gamma}_{1}-(\beta_{0}\,\beta_{2}-\beta_{1}^{2})\,\overline{\gamma}_{0}\right)}{2\beta_{0}^{3}} \\
     &+ h^{3}\,\bigl(\beta_{0}^{3}\,\overline{\gamma}_{3}-\beta_{0}^{2}\,\beta_{1}\,\overline{\gamma}_{2}+(\beta_{0}\,\beta_{1}^{2}-\beta_{0}^{2}\,\beta_{2})\,\overline{\gamma}_{1} \\
     &\qquad +(-\beta_{0}^{2}\,\beta_{3}+2\,\beta_{0}\,\beta_{1}\,\beta_{2}-\beta_{1}^{3})\,\overline{\gamma}_{0}\bigr)\frac{1}{3\beta_{0}^{4}}
     \end{split}
\end{align}
with the standard four-loop formula for the running coupling
\begin{align}
  \label{eq:alpha}
  \begin{split}
      h(t) &= \frac{1}{\beta_{0}t}
      \left(1 - \frac{\beta_{1}}{\beta_{0}^{2}}\frac{\log(t)}{t}
     + \frac{\beta_{1}^{2}}{\beta_{0}^{4}}
       \frac{1}{t^{2}}\left(\left(\log(t)-\frac{1}{2}\right)^{2}
     + \frac{\beta_{2}\beta_{0}}{\beta_{1}^{2}}-\frac{5}{4}\right)\right) \\
     &+ \frac{1}{(\beta_{0}t)^{4}}
 \left(\frac{\beta_{3}}{2\beta_{0}}+
   \frac{1}{2}\left(\frac{\beta_{1}}{\beta_{0}}\right)^{3}
   \left(-2\log^{3}(t)+5\log^{2}(t)+
\left(4-6\frac{\beta_{2}\beta_{0}}{\beta_{1}^{2}}\right)\log(t)-1\right)\right)
     \end{split}
\end{align}
and $\displaystyle{t=\log\left(\frac{\mu^{2}}{\Lambda^{2}}\right)}$.

\begin{figure}[h]
  \centering
  \psfig{figure=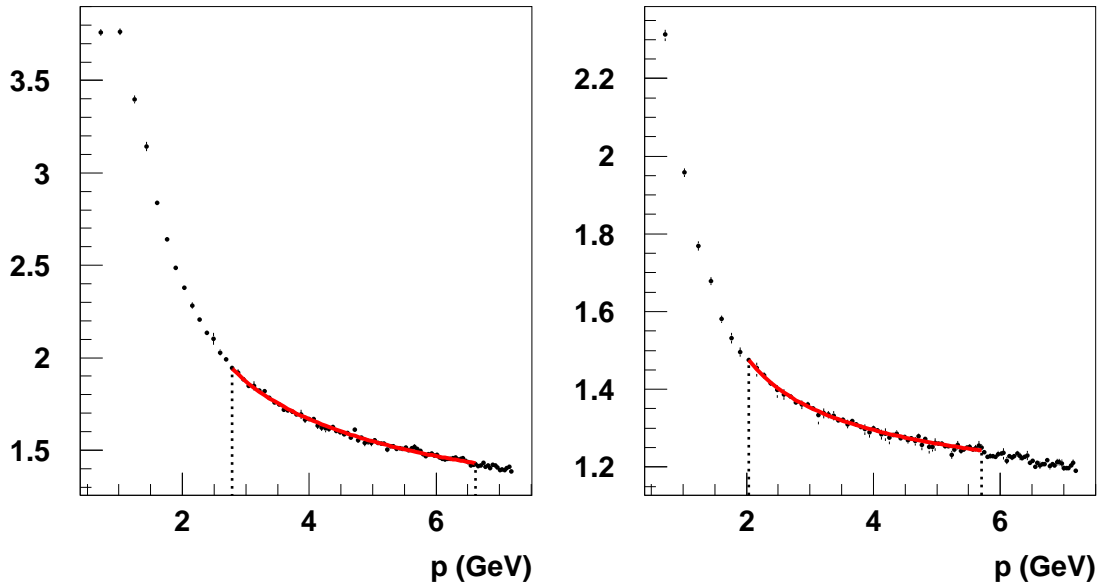, width=17cm, height=9cm}
  \caption{Extrapolated lattice data at $\beta=6.4$ for $Z_{3}$ (left)
    and $\widetilde{Z}_{3}$ (right). The solid line is the fit at
    four-loop order in the $\ms$ scheme. The vertical dotted lines
    delimit the window of each fit.}
  \label{fig:Zn}
\end{figure}

We now consider in turn the three renormalization schemes $\ms$,
$\momg$ and $\momc$ and fit the two parameters of
Eqs.\,(\ref{eq:Zmom}) and (\ref{eq:alpha}) to our extrapolated lattice
data. Figure\,\ref{fig:Zn} illustrates the typical quality of such fits.

\subsection{$\ms$ scheme}
\label{sec:ms}

The analysis in the $\ms$ scheme is summarized in Table~\ref{tab:ms}.
The statistical error is at the level of 1\% for the gluon propagator
and 2-3\% for the ghost propagator, whereas the systematic error due
to the extrapolations is around 3-5\% and 5-10\% respectively. The
values of $\Lambda_{\ms}$ extracted from the gluon and the ghost
propagators are consistent within these errors and within each order
of perturbation theory.

\begin{table}[h]
\centering
\begin{tabular}{c|c|c|c|c|c|c|c|c}
\hline
\hline
$\beta$ & $L$ & $a$p$_{min}$ & $a\Lambda^{(2)}_{\ms,gluon}$ & $\chi^{2}$
              & $a\Lambda^{(3)}_{\ms,gluon}$ & $\chi^{2}$ 
              & $a\Lambda^{(4)}_{\ms,gluon}$ & $\chi^{2}$ \\
\hline
6.0 & 16 & 1.111 & 0.336(3)$^{+8~}_{-4~}$ & 1.3  
                 & 0.265(3)$^{+6~}_{-2~}$ & 1.0
                 & 0.225(2)$^{+4~}_{-2~}$ & 1.1  \\
\hline
    & 24 & 1.111 & 0.332(3)$^{+8~}_{-12}$ & 0.6 
                 & 0.262(3)$^{+6~}_{-8~}$ & 0.5 
                 & 0.222(2)$^{+5~}_{-8~}$ & 0.6  \\
\hline
6.2 & 24 & 0.907 & 0.240(2)$^{+6~}_{-9~}$  & 0.8
                 & 0.185(2)$^{+6~}_{-10}$  & 0.8 
                 & 0.158(2)$^{+4~}_{-7~}$  & 0.8  \\
\hline
6.4 & 32 & 0.760 & 0.171(2)$^{+10}_{-11}$ & 1.4
                 & 0.130(2)$^{+12}_{-11}$ & 1.4 
                 & 0.112(1)$^{+9~}_{-8~}$  & 1.4 \\
\hline
\hline
$\beta$ & $L$ & $a$p$_{min}$ & $a\Lambda^{(2)}_{\ms,ghost}$ & $\chi^{2}$
              & $a\Lambda^{(3)}_{\ms,ghost}$ & $\chi^{2}$ 
              & $a\Lambda^{(4)}_{\ms,ghost}$ & $\chi^{2}$ \\
\hline
6.0 & 16 & 1.039 & 0.354(7)$^{+23}_{-13}$ & 0.5  
                 & 0.281(6)$^{+17}_{-8~}$ & 0.5
                 & 0.235(5)$^{+15}_{-7~}$ & 0.5  \\
\hline
    & 24 & 0.785 & 0.325(6)$^{+13}_{-20}$ & 0.2 
                 & 0.259(5)$^{+10}_{-18}$ & 0.2 
                 & 0.217(4)$^{+8~}_{-13}$ & 0.2  \\
\hline
6.2 & 24 & 0.693 & 0.254(4)$^{+20}_{-20}$  & 0.4
                 & 0.200(3)$^{+10}_{-23}$  & 0.4 
                 & 0.169(3)$^{+12}_{-13}$  & 0.4  \\
\hline
6.4 & 32 & 0.555 & 0.193(2)$^{+22}_{-14}$ & 0.8
                 & 0.150(2)$^{+15}_{-14}$ & 0.8 
                 & 0.128(2)$^{+13}_{-11}$  & 0.8 \\
\hline
\hline
\end{tabular}
\caption{Fits of $\Lambda_{\ms}$ from the gluon and ghost lattice propagators.
The error in parenthesis is the statistical error corresponding to a window
$[a\,p_{min},a\,p_{max}]$ with the $a\,p_{min}$ quoted in the Table and 
the upper bound for $a\,p_{max}$ quoted in Tables~\ref{tab:cutg} and 
\ref{tab:cutc} respectively.}
\label{tab:ms}
\end{table}

 However, the three-loop and four-loop values, which are displayed in
Table~\ref{tab:ms_values} with the physical units of
Table~\ref{tab:simulation}, clearly confirm our previous result
\cite{LPTX99} that we are still far from asymptoticity in that scheme.

\begin{table}[h]
\centering
\begin{tabular}{c||c|c||c|c}
\hline
$\beta$ & $\Lambda^{(3)}_{\ms,gluon}$ & $\Lambda^{(3)}_{\ms,ghost}$ &
$\Lambda^{(4)}_{\ms,gluon}$ & $\Lambda^{(4)}_{\ms,ghost}$ \\
\hline
6.0 &  $519(6)^{+12}_{-4~}$ & $551(12)^{+33}_{-16}$ & $441(4)^{+8~}_{-4~}$ & 
$461(10)^{+29}_{-14}$ \\
6.2 &  $509(6)^{+17}_{-27}$ & $550(8)^{+27}_{-63}~$ & $435(6)^{+11}_{-19}$ & 
$465(8)^{+33}_{-36}~$ \\
6.4 &  $476(7)^{+44}_{-40}$ & $549(7)^{+55}_{-51}~$ & $410(4)^{+33}_{-29}$ & 
$468(7)^{+48}_{-40}~$ \\
\hline
\end{tabular}
\caption{Three-loop and four-loop physical values of $\Lambda_{\ms}$ in MeV
  extracted from Table~\ref{tab:ms}.}
\label{tab:ms_values}
\end{table}

\subsection{$\momg$ scheme}
\label{sec:momg}

Table~\ref{tab:momg}, which summarizes the analysis in the $\momg$
scheme, shows that, at the lower $\beta$'s, we were not able to
describe both lattice propagators at four-loop order with reasonable
cuts and $\chi^{2}$. This could be interpreted as an hint that
perturbation theory has some problems of convergence beyond three-loop
order below 3-4 GeV.

\begin{table}[h]
\centering
\begin{tabular}{c|c|c|c|c|c|c|c|c}
\hline
\hline
$\beta$ & $L$ & $a$p$_{min}$ & $a\Lambda^{(2)}_{\momg,gluon}$ & $\chi^{2}$
              & $a\Lambda^{(3)}_{\momg,gluon}$ & $\chi^{2}$ 
              & $a\Lambda^{(4)}_{\momg,gluon}$ & $\chi^{2}$ \\
\hline
6.0 & 16 & 1.039 & 0.551(3)$^{+8~}_{-8~}$ & 1.0  
                 & 0.477(3)$^{+5~}_{-8~}$ & 1.2
                 & --- & ---  \\
\hline
    & 24 & 1.014 & 0.536(4)$^{+14}_{-19}$ & 0.9 
                 & 0.464(3)$^{+10}_{-11}$ & 0.9 
                 & --- & ---  \\
\hline
6.2 & 24 & 0.693 & 0.396(2)$^{+19}_{-12}$ & 1.0
                 & 0.336(2)$^{+8~}_{-15}$  & 0.9 
                 & --- & --- \\
\hline
6.4 & 32 & 0.555 & 0.292(1)$^{+15}_{-14}$ & 1.3
                 & 0.246(1)$^{+7~}_{-20}$ & 1.4 
                 & 0.253(3)$^{+5~}_{-3~}$ & 1.6 \\
\hline
\hline
$\beta$ & $L$ & $a$p$_{min}$ & $a\Lambda^{(2)}_{\momg,ghost}$ & $\chi^{2}$
              & $a\Lambda^{(3)}_{\momg,ghost}$ & $\chi^{2}$ 
              & $a\Lambda^{(4)}_{\momg,ghost}$ & $\chi^{2}$ \\
\hline
6.0 & 16 & 1.039 & 0.660(40)$^{+24}_{-29}$ & 0.4  
                 & 0.475(12)$^{+29}_{-24}$ & 0.5
                 & --- & --- \\
\hline
    & 24 & 1.014 & 0.559(22)$^{+25}_{-20}$ & 0.2 
                 & 0.438(12)$^{+14}_{-25}$ & 0.2
                 & 0.408(17)$^{+20}_{-18}$ & 0.9  \\
\hline
6.2 & 24 & 0.693 & 0.455(11)$^{+9~}_{-17}$ & 0.3
                 & 0.342(5)$^{+27}_{-34}~$ & 0.6 
                 & 0.348(8)$^{+23}_{-17}~$ & 1.0 \\
\hline
6.4 & 32 & 0.555 & 0.333(4)$^{+36}_{-26}~$ & 1.2
                 & 0.261(3)$^{+33}_{-28}~$ & 0.9 
                 & 0.279(7)$^{+14}_{-30}~$ & 0.8 \\
\hline
\hline
\end{tabular}
\caption{Fits of $\Lambda_{\momg}$ from the gluon and ghost lattice
propagators. The notations are the same as in Table~\ref{tab:ms}.}
\label{tab:momg}
\end{table}

If we select the three-loop result as the best perturbative estimate
of $\Lambda_{\momg}$ and convert it to the $\ms$ scheme with the
asymptotic one-loop formula, $\Lambda_{\ms} = 0.346\,\Lambda_{\momg}$,
then we get the physical values quoted in Table~\ref{tab:momg_values}
which agree completely with previous values \cite{LPTX00}.

\begin{table}[h]
\centering
\begin{tabular}{c||c|c}
\hline
$\beta$ & $\Lambda^{(3)}_{\ms,gluon}$ & $\Lambda^{(3)}_{\ms,ghost}$ \\
\hline
6.0 &  324(2)$^{+2~}_{-5~}$ & 322(8)$^{+20}_{-16}$  \\
6.2 &  320(2)$^{+8~}_{-14}$ & 326(5)$^{+26}_{-33}$   \\
6.4 &  312(1)$^{+9~}_{-25}$ & 331(4)$^{+42}_{-35}$  \\
\hline
\end{tabular}
\caption{Three-loop physical values of $\Lambda_{\ms}$ in MeV
  extracted from Table~\ref{tab:momg}.}
\label{tab:momg_values}
\end{table}

\subsection{$\momc$ scheme}
\label{sec:momc}

The results of the analysis in the $\momc$ scheme are displayed in
Table~\ref{tab:momc}. We still find that the three-loop and four-loop
values of $\Lambda^{(3)}_{\momc}$ are very much the same both for the
gluon propagator and for the ghost propagator. Thus the perturbative
series seems again to become asymptotic at three-loop order in that
scheme.

\begin{table}[h]
\centering
\begin{tabular}{c|c|c|c|c|c|c|c|c}
\hline
\hline
$\beta$ & $L$ & $a$p$_{min}$ & $a\Lambda^{(2)}_{\momc,gluon}$ & $\chi^{2}$
              & $a\Lambda^{(3)}_{\momc,gluon}$ & $\chi^{2}$ 
              & $a\Lambda^{(4)}_{\momc,gluon}$ & $\chi^{2}$ \\
\hline
6.0 & 16 & 1.178 & 0.482(6)~~~~ & 1.3  
                 & 0.408(3)$^{+4~}_{-4~}~$ & 1.0
                 & --- & --- \\
\hline
    & 24 & 1.111 & 0.468(5)$^{+6~}_{-5~}~$ & 0.5 
                 & 0.394(3)$^{+11}_{-6~}~$ & 0.8 
                 & 0.411(7)$^{+3~}_{-2~}~$ & 1.1  \\
\hline
6.2 & 24 & 0.907 & 0.345(3)$^{+17}_{-10}~$ & 0.9
                 & 0.288(2)$^{+5~}_{-6~}~$ & 0.9 
                 & 0.292(3)$^{+8~}_{-5~}~$ & 0.8 \\
\hline
6.4 & 32 & 0.589 & 0.255(1)$^{+12}_{-15}~$ & 1.5
                 & 0.205(1)$^{+11}_{-7~}~$ & 1.6 
                 & 0.212(1)$^{+9~}_{-17}~$ & 1.6 \\
\hline
\hline
$\beta$ & $L$ & $a$p$_{min}$ & $a\Lambda^{(2)}_{\momc,ghost}$ & $\chi^{2}$
              & $a\Lambda^{(3)}_{\momc,ghost}$ & $\chi^{2}$ 
              & $a\Lambda^{(4)}_{\momc,ghost}$ & $\chi^{2}$ \\
\hline
6.0 & 16 & 0.962 & 0.489(6)$^{+10}_{-6~}~$ & 0.8  
                 & 0.437(11)$^{+6~}_{-2~}$ & 0.4
                 & --- & ---  \\
\hline
    & 24 & 1.047 & 0.459(6)$^{+15}_{-15}~$ & 0.5
                 & 0.408(8)$^{+6~}_{-5~}~$ & 0.5 
                 & 0.398(14)$^{+13}_{-20}$ & 0.3  \\
\hline
6.2 & 24 & 0.740 & 0.367(7)$^{+21}_{-33}~$ & 0.4
                 & 0.308(7)$^{+9~}_{-16}~$ & 0.2 
                 & 0.303(9)$^{+7~}_{-14}~$ & 0.2  \\
\hline
6.4 & 32 & 0.589 & 0.280(5)$^{+28}_{-23}~$ & 0.6
                 & 0.225(5)$^{+18}_{-13}~$ & 0.6 
                 & 0.224(5)$^{+15}_{-16}~$ & 0.6 \\
\hline
\hline
\end{tabular}
\caption{Fits of $\Lambda_{\momc}$ from the gluon and ghost lattice
propagators. The notations are the same as in Table~\ref{tab:ms}.}
\label{tab:momc}
\end{table}

Selecting the three-loop result as the best perturbative estimate of
$\Lambda_{\momc}$ and converting it to the $\ms$ scheme with the
asymptotic formula, $\Lambda_{\ms} = 0.429\,\Lambda_{\momc}$, we get the
physical values quoted in Table~\ref{tab:momc_values}.

\begin{table}[h]
\centering
\begin{tabular}{c||c|c}
\hline
$\beta$ & $\Lambda^{(3)}_{\ms,gluon}$ & $\Lambda^{(3)}_{\ms,ghost}$ \\
\hline
6.0 &  345(3)$^{+4~}_{-4~}$ & 369(9)$^{+3~}_{-2~}$  \\
6.2 &  341(2)$^{+6~}_{-7~}$ & 364(8)$^{+11}_{-19}$  \\
6.4 &  323(2)$^{+17}_{-11}$ & 354(8)$^{+28}_{-20}$  \\
\hline
\end{tabular}
\caption{Three-loop physical values of $\Lambda_{\ms}$ in MeV
  extracted from Table~\ref{tab:momc}.}
\label{tab:momc_values}
\end{table}

\subsection{Scheme dependence}

The puzzling feature of Tables~\ref{tab:ms_values},
\ref{tab:momg_values} and \ref{tab:momc_values} is the rather large
dependence of the $\Lambda_{\text{QCD}}$ scale upon the loop-order and
the renormalisation scheme whereas, within any scheme, the values from
the ghost and gluon propagators are rather consistent at each loop
order and pretty independent of the lattice spacing.

Let us consider again the evolution equation of the renormalisation
constants of the gluon or ghost fields in a MOM scheme, with respect
to the coupling $h_{R}$ in an arbitrary scheme $R$. We have
\begin{align}
  \label{eq:dlambda}
  \frac{d\ln(Z_{\Gamma,MOM})}{d \ln \mu^{2}} = \overline{\gamma}_{R}(h_{R}) =
  -\frac{1}{2} \frac{d\ln(Z_{\Gamma,MOM})}{d \ln \Lambda_{R}}
\end{align}
where $ \Lambda_{R}$ is the scale in scheme $R$. If we truncate the
perturbative expansion at order $n$
\begin{align}
  \label{eq:Zn}
  \ln\left(\frac{Z_{\Gamma,MOM}}{Z_{0}}\right) = 
  c_{R,0}\ln(h_{R}) + \sum_{k=1}^{n-1} c_{R,k}h_{R}^{k}
\end{align}
the change in the effective scale $\Lambda^{(n)}_{R}$, or
equivalently, the change in the coupling $h_{R}$, induced by adding
the contribution at order $n+1$ is typically
\begin{align}
  \label{eq:relative}
  \frac{\Delta\Lambda^{(n)}_{R}}{\Lambda^{(n)}_{R}} \approx 
  -\frac{c_{R,n}h_{R}^{n}}{2\gamma_{R}(h_{R})}
\end{align}
Now the dependence of the effective scale $\Lambda_{R}$ upon the coupling
$h_{R}$ is given up to order 4
\begin{align}
  \label{eq:beta}
  2\ln\Lambda^{(4)}_{R} = \ln\mu^{2} - \frac{1}{\beta_{0}h_{R}} - 
  \frac{\beta_{1}}{\beta_{0}^{2}}\ln(\beta_{0}h_{R}) - 
  \frac{\beta_{0}\beta_{2}-\beta_{1}^{2}}{\beta_{0}^{3}}h_{R} -
  \frac{\beta_{0}^{2}\beta_{3} - 2\beta_{0}\beta_{1}\beta_{2} +
    \beta_{1}^{3}}{2\beta_{0}^{4}}h_{R}^{2}
\end{align}
and, denoting the coefficient of order $h_{R}^{n-2}$ in that equation by
$-\rho_{R,n-1}$, the effective scales which describe a same coupling at
order $n$ and $n+1$ are related by
\begin{align}
  \label{eq:order}
  \ln\frac{\Lambda^{(n+1)}_{R}}{\Lambda^{(n)}_{R}} \equiv 
  -\frac{1}{2} \rho_{R,n-1}h_{R}^{n-1}
\end{align}
Combining Eqs.\,(\ref{eq:relative}) and (\ref{eq:order}) gives the
relation between the effective scales which describe the
renormalisation constants of the gluon or ghost fields in a MOM
scheme at order $n$ and $n+1$
\begin{align}
  \label{eq:loop}
  \frac{\Lambda^{(n+1)}_{R}}{\Lambda^{(n)}_{R}} = \exp-\frac{1}{2}
  \left(\rho_{R,n-1}+\frac{c_{R,n}h_{R}}{\gamma_{R}(h_{R})}\right)h_{R}^{n-1}
\end{align}

\begin{figure}[h]
  \centering
  \psfig{figure=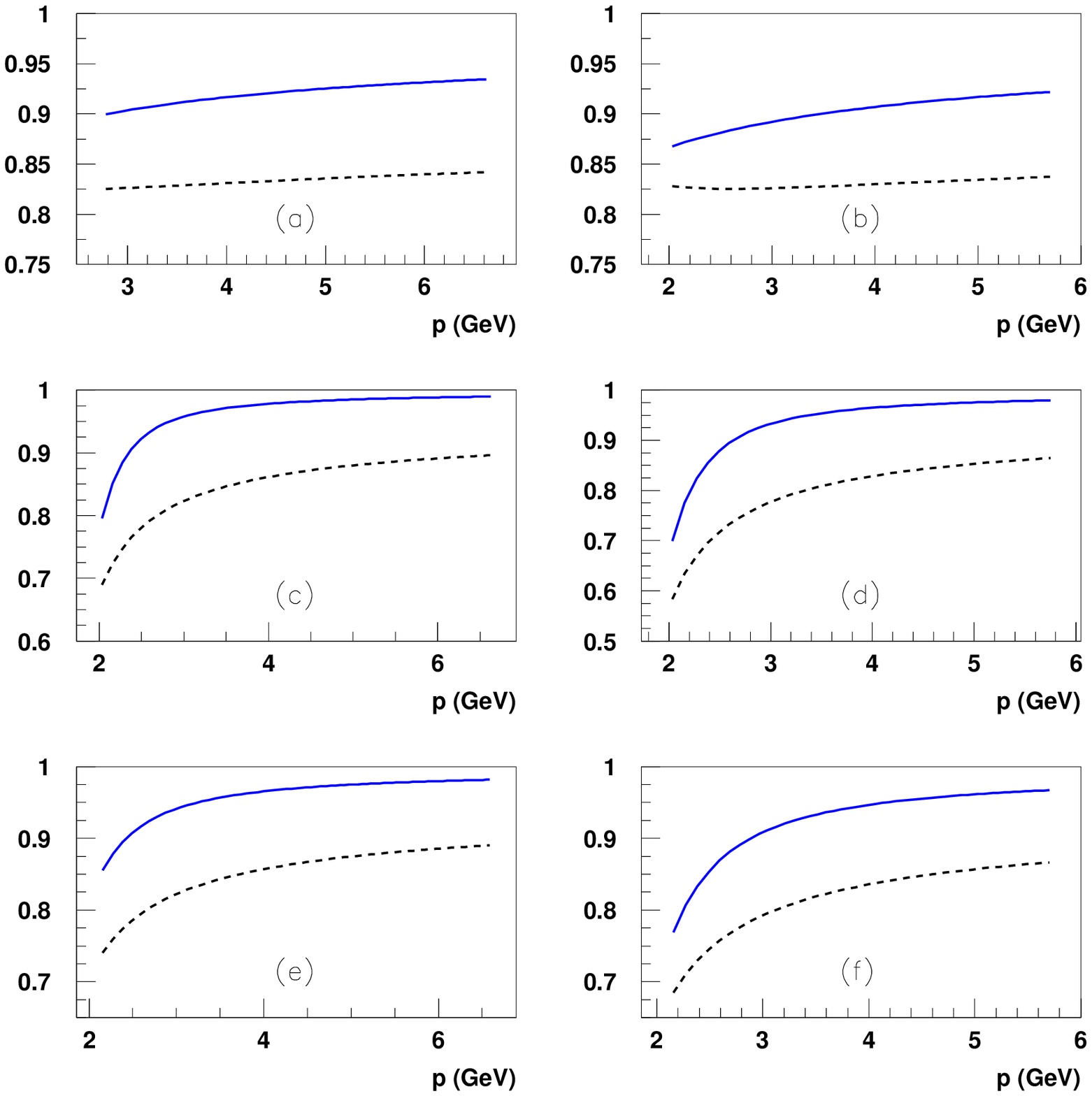, width=15cm, height=16cm}
  \caption{ $\frac{\Lambda^{(n+1)}_{R}}{\Lambda^{(n)}_{R}}$ for
    $n=2$ (dashed lines) and $n=3$ (solid lines), for the
    gluon propagator in the $\ms$ scheme (a), $\momg$ scheme (c) and
    $\momc$ scheme (e), and for the ghost propagator in the $\ms$
    scheme (b), $\momg$ scheme (d) and $\momc$ scheme (f).}
  \label{fig:errors}
\end{figure}

Figure~\ref{fig:errors} displays the behavior of this ratio for the
gluon and ghost propagators in the three schemes as a function of the
momentum $p$ for $n=2$ and $n=3$. The couplings are taken from
the fits at $\beta=6.4$.
There is a pretty good qualitative agreement with
Tables~\ref{tab:ms_values}, \ref{tab:momg_values} and
\ref{tab:momc_values}, which confirms the overall consistency with
perturbation theory of the lattice data for the gluon and ghost
propagators \bfit{within} any renormalization scheme.

\begin{figure}[h]
  \centering
  \psfig{figure=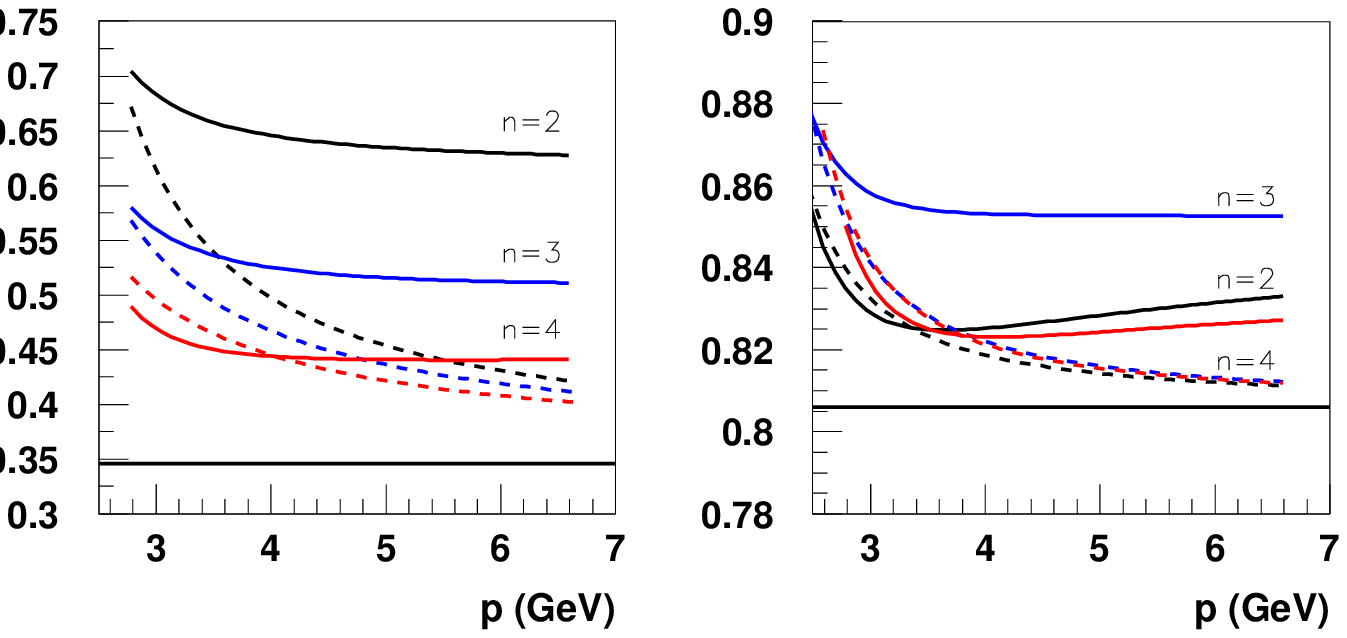, width=16cm, height=9cm}
  \caption{$\frac{\Lambda^{(n)}_{\ms}}{\Lambda^{(n)}_{\momg}}$ (left)
    and $\frac{\Lambda^{(n)}_{\momc}}{\Lambda^{(n)}_{\momg}}$ (right)
    for $n=2$, $n=3$ and $n=4$. The solid lines are the plots of
    Eq.\,(\ref{eq:scheme}) with the fitted couplings whereas the
    dashed lines are the plots with Eq.\,(\ref{eq:expand}). Horizontal
    lines are the asymptotic values.}
  \label{fig:scheme}
\end{figure}

The scheme dependence of the $\Lambda_{\text{QCD}}$ scale can also be
analyzed with Eq.\,(\ref{eq:beta}):
\begin{align}
  \label{eq:scheme}
  \frac{\Lambda^{(n)}_{R_{2}}}{\Lambda^{(n)}_{R_{1}}} = \exp\left\{
  \frac{1}{2\beta_{0}}\left(\frac{1}{h_{R_{1}}}-\frac{1}{h_{R_{2}}}\right) +
  \frac{\beta_{1}}{2\beta_{0}^{2}}\ln\frac{h_{R_{1}}}{h_{R_{2}}} +
  \cdots \right\}
\end{align}
Figure~\ref{fig:scheme} shows the behavior of the ratios
$\frac{\Lambda^{(n)}_{\ms}}{\Lambda^{(n)}_{\momg}}$ and
$\frac{\Lambda^{(n)}_{\momc}}{\Lambda^{(n)}_{\momg}}$, as a function of
the momentum $p$ at each order of perturbation theory. The couplings
are taken from the fits of the gluon propagator at $\beta=6.4$.

\begin{figure}[h]
  \centering
  \psfig{figure=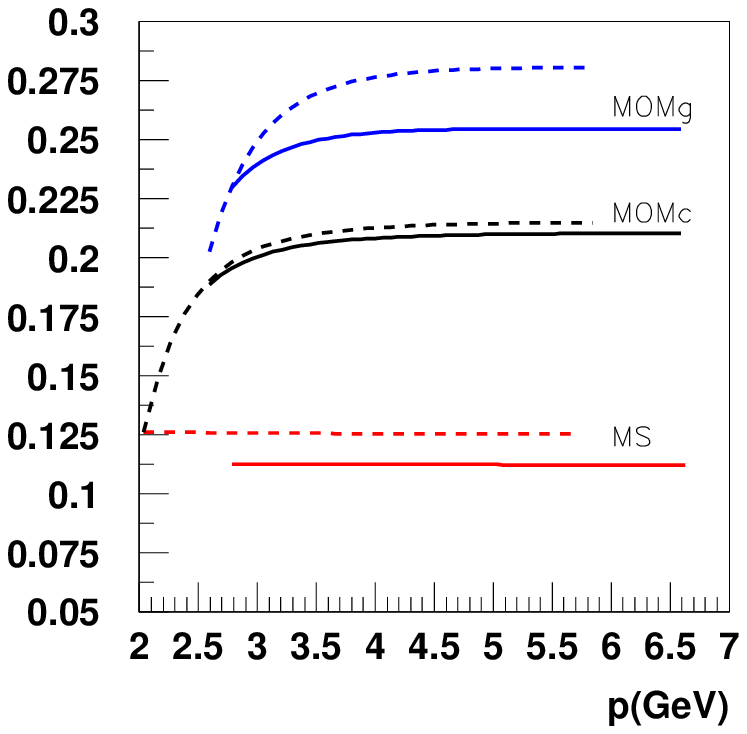, width=10cm, height=10cm}
  \caption{$a\Lambda^{(4)}_{\ms}$, $a\Lambda^{(4)}_{\momg}$ and
    $a\Lambda^{(4)}_{\momc}$ from the gluon propagator (solid lines) and
    from the ghost propagator (dashed lines) at $\beta$ = 6.4, as a
    function of the momentum through Eq.\,(\ref{eq:beta}).}
  \label{fig:lambda}
\end{figure}

Clearly, the limiting values of these ratios are not the asymptotic
values. If we replace in Eq.\,(\ref{eq:scheme}) the coupling
$h_{R_{2}}$ by its perturbative expansion with respect to $h_{R_{1}}$
\begin{align}
  h_{R_{2}} = h_{R_{1}} + \sum_{k=1}^{n-1} r_{k} h_{R_{1}}^{k+1}
  \label{eq:expand}
\end{align}
then the ratios do of course tend towards the asymptotic values
$\displaystyle{\exp\left\{\frac{r_{1}}{2\beta_{0}}\right\}}$. The
disagreement with respect to the perturbative expansion is not a
problem with the lattice data or with the numerical analysis. Indeed
the fits do a very good job at extracting a well-behave coupling as
illustrated in Fig.~\ref{fig:lambda} which displays the dimensionless
scales $a\Lambda^{(4)}_{\ms}$, $a\Lambda^{(4)}_{\momg}$ and
$a\Lambda^{(4)}_{\momc}$ as a function of the momentum $p$, using
Eq.\,(\ref{eq:beta}) with the fitted couplings at $\beta = 6.4$ from
the ghost and gluon propagators.  $Z_{0}$, the other fitted parameter
of Eq.\,(\ref{eq:Zmom}), is nearly independent, within a few percent,
of the renormalisation scheme as it should in the absence of
truncations. It follows that the difficulty to reproduce the
asymptotic ratios between the scales of different renormalization
schemes, is mainly a consequence of the truncation of the perturbative
series of the renormalization constants of the gluon and ghost
propagators.

\begin{figure}[h]
  \centering
  \psfig{figure=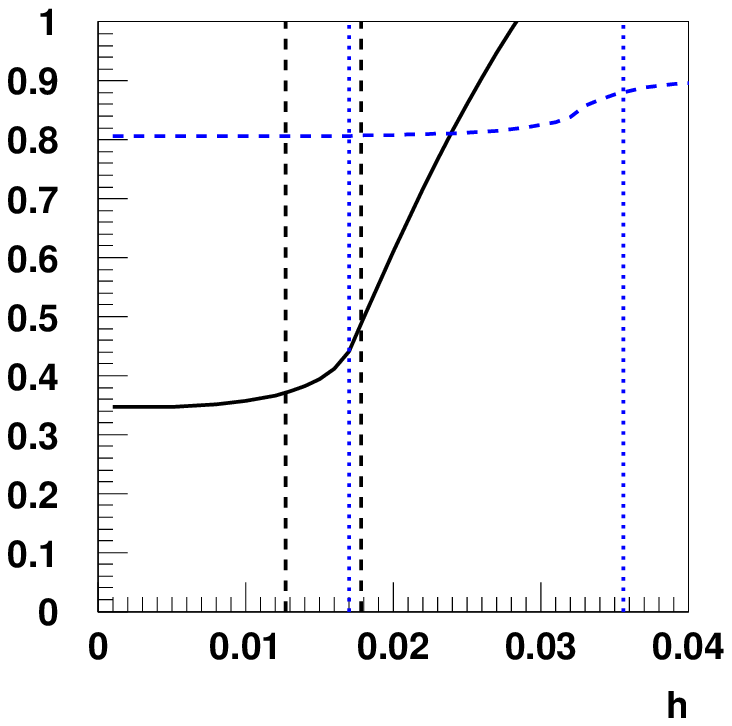, width=10cm, height=10cm}
  \caption{$\frac{\Lambda^{(4)}_{\ms}}{\Lambda^{(4)}_{\momg}}$ (solid
    line) as a function of $h_{\ms}$ and
    $\frac{\Lambda^{(4)}_{\momc}}{\Lambda^{(4)}_{\momg}}$ (dashed
    line) as a function of $h_{\momc}$. The vertical lines delimit the
    values spanned by $h_{\ms}$ (dashed) and $h_{\momc}$ (dotted)
    in the fits of the gluon propagator at $\beta=6.4$.}
  \label{fig:truncation}
\end{figure}

We can substantiate this claim, and estimate the rate of convergence,
by the following exercise. We solve $h_{R_{2}}$ in terms of
$h_{R_{1}}$ using Eq.\,(\ref{eq:Zn}) at four-loop order
\begin{align} 
  \ln\left(\frac{Z_{\Gamma,MOM}}{Z_{0}}\right) = 
  c_{R_{2},0}\ln(h_{R_{2}}) + \sum_{k=1}^{3} c_{R_{2},k}h_{R_{2}}^{k} = 
  c_{R_{1},0}\ln(h_{R_{1}}) + \sum_{k=1}^{3} c_{R_{1},k}h_{R_{1}}^{k}
\end{align}
Then we plug the solution into Eq.\,(\ref{eq:scheme}).
Figure~\ref{fig:truncation} shows the behavior of the corresponding
ratios, $\frac{\Lambda^{(4)}_{\ms}}{\Lambda^{(4)}_{\momg}}$ and
$\frac{\Lambda^{(4)}_{\momc}}{\Lambda^{(4)}_{\momg}}$, as a function
of the coupling $h_{\ms}$ and $h_{\momc}$ respectively. The effect of
the truncation of the perturbative series is manifest for the $\ms$
scheme and gives the right order of magnitude of what is actually
measured in Tables~\ref{tab:ms} and \ref{tab:momg}.

\section{Conclusion}
\label{sec:conclusion}

We have shown that the lattice formulation of the ghost propagator has
the expected perturbative behavior up to four-loop order from 2 GeV to
6 GeV.  We have been able to go beyond the qualitative level and to
produce quantitative results for the scale $\Lambda_{\ms}$ which are
pretty consistent with the values extracted from the lattice gluon
propagator. We have understood the strong dependence of the effective
$\Lambda_{\ms}$ scale upon the order of perturbation theory and upon
the renormalisation scheme used for the parametrisation of the data.
The perturbative series of the $\mom$ schemes seem to be asymptotic at
three-loop order in the energy range we have probed whereas the $\ms$
scheme converges very slowly. If we assume that all perturbative
series remain well behaved beyond four-loop above 4 GeV, then we get
$\Lambda_{\ms} \approx 320$ MeV with a 10\% systematic uncertainty.
The statistical errors are at the 1\% level. This value is also in
pretty good agreement with the values of $\Lambda_{\ms}$ extracted
from the three-gluon vertex in a $\mom$ scheme at three-loop order
\cite{LPTX98}, at the same $\beta$'s and with the same lattice sizes.
On the other hand it exceeds by 20\% the value obtained from the same
vertex at $\beta=6.8$ on a $24^{4}$ lattice.  This discrepancy
motivated the introduction of power corrections which are successful
in describing the combined data of the three-gluon vertex
\cite{BOU00}. We will show in a forthcoming paper how the power
corrections can be unraveled from the lattice propagators alone.

\begin{figure}[h]
  \centering
  \psfig{figure=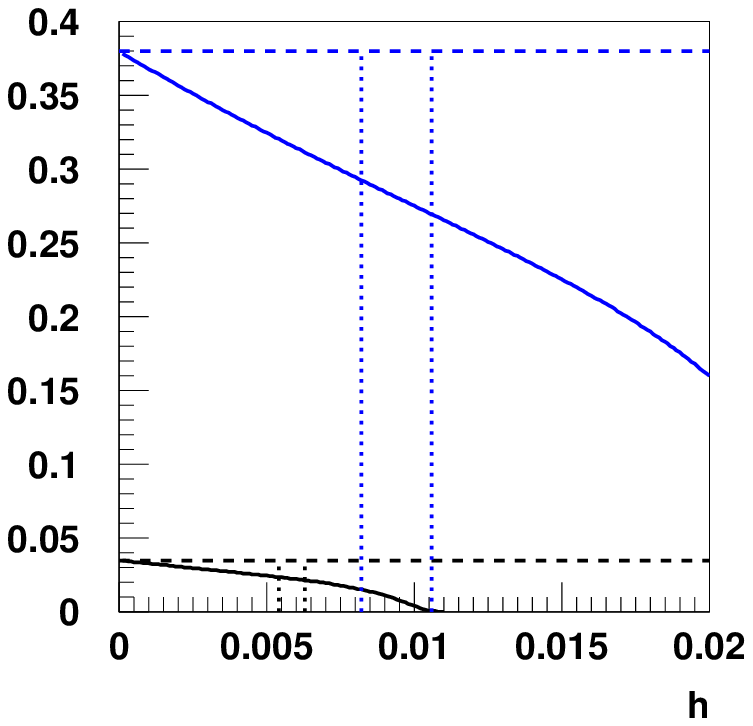, width=10cm, height=10cm}
  \caption{$\frac{\Lambda^{(3)}_{\text{L}}}{\Lambda^{(3)}_{\ms}}$
    (lower solid line) as a function of $h_{\text{L}}$ and
    $\frac{\Lambda^{(3)}_{\square}}{\Lambda^{(3)}_{\ms}}$ (upper solid
    line) as a function of $h_{\square}$. The vertical lines
    (dotted) delimit the values spanned by $h_{\text{L}}$ and
    $h_{\square}$ in the simulations of \cite{GOC05} $(5.7 \leq \beta
    \leq 6.9)$. The dashed horizontal lines are the asymptotic
    values.}
  \label{fig:lattice}
\end{figure}

The value quoted above exceeds also by about 30\% the previous
determinations of the QCD scale in the quenched approximation based on
gauge-invariant definitions of the strong coupling constant
\cite{ALPHA,BOO01} (take note, for comparison purposes, that our
physical unit corresponds to the force parameter $r_{0}$ \cite{NEC02}
set approximately to 0.53\,fm). However there is also an uncertainty
due to the use of the asymptotic one-loop relation between
$\Lambda_{\ms}$ and the $\Lambda_{\text{L}}$'s. For illustration, let
us consider the determination of $\Lambda_{\ms}$ using lattice
perturbation theory up to three-loop order with the Wilson action
\cite{GOC05}. It is possible to estimate the rate of convergence of
the ratio $\frac{\Lambda^{(3)}_{\text{L}}}{\Lambda^{(3)}_{\ms}}$ as a
function of the bare lattice coupling $h_{\text{L}} =
\frac{6}{(4\pi)^{2}\beta}$ by inserting the perturbative expansion of
$h_{\ms}$ into Eq.\,(\ref{eq:scheme}). Figure~\ref{fig:lattice}
displays the evolution of this ratio and also of the ratio
$\frac{\Lambda^{(3)}_{\square}}{\Lambda^{(3)}_{\ms}}$ for the
so-called ``boosted'' lattice scheme which re-express the lattice
perturbative series as a function of the coupling
$h_{\square}=h_{\text{L}}/\left<plaq\right>$.
The mere truncation of the perturbative series introduces an
uncertainty on the absolute scale of the lattice schemes which could
be as large as 30\% in the range of $\beta$ studied in these
simulations.

No strategy can fix the scale $\Lambda_{\text{QCD}}$ to an accuracy
better than the uncertainty entailed by the truncation of the
perturbative series in the conversion to the $\ms$ scheme. We have
shown that this error can be larger than the main well-known sources
of systematic errors which come from setting the scale $a^{-1}$ and
from the continuum extrapolation. If we aim at reducing below 10\% the
error in the conversion of the $\mom$ schemes to the $\ms$ scheme,
then a look at Figure~\ref{fig:truncation} shows that we need to apply
a cut at 6 GeV. Such an analysis would require simulations at
$\beta=6.6$ and $\beta=6.8$ on $48^{4}$ and $64^{4}$ lattices
respectively, to work at fixed volume and minimize finite-size
effects.  The existence of several lattice observables, gluon
propagator, ghost propagator, three-gluon vertex, from which one can
extract independent values of the scale $\Lambda_{\text{QCD}}$, an
advantage of the Green function approach, should then allow to
disentangle unambiguously the effects of the truncation of the
perturbative series from the non-perturbative corrections, and to get
a value of $\Lambda_{\ms}$ at a true 10\% accuracy.

\end{document}